\documentclass[12pt]{article}
\pdfoutput=1
\usepackage{amsmath,amssymb,epsfig,epic,amscd}
\usepackage{color}
\usepackage{graphicx}
\usepackage{epstopdf}

\begin{document}
\title{Weak Lax pairs for  lattice equations}
\author{Jarmo Hietarinta$^{1,2}$\footnote{E-mail:
    jarmo.hietarinta@utu.fi}~ and Claude Viallet$^{1}\footnote{E-mail:
    claude.viallet@upmc.fr}~$ \\ {\small\it $^1$ LPTHE / CNRS / UPMC,
    4 place Jussieu 75252 Paris CEDEX 05, France }\\ {\small\it
    $^2$Department of Physics and Astronomy, University of Turku,
    FIN-20014 Turku, Finland}}

\maketitle

\begin{abstract}
We consider various 2D lattice equations and their integrability, from
the point of view of 3D consistency, Lax pairs and B\"acklund
transformations.  We show that these concepts, which are associated
with integrability, are not strictly equivalent. In the course of our
analysis, we introduce a number of black and white lattice models, as
well as variants of the functional Yang-Baxter equation.
\end{abstract}

\section{Introduction}
Recent progress in the description of integrable partial difference
equations is to a great part due to the consistency approach
\cite{NRGO,NW,BS-IMR,ABS}, in particular in the form of 3 dimensional
Consistency-Around-a-Cube (CAC). One of the highlights of this
approach is the immediate existence of a Lax pair and B\"acklund
transforms (BT), which can be directly constructed from the
``side-equations'' of the cube \cite{NW,Nlax}. One can say, in effect,
that the set of side equations yields the BT and the Lax pair. One
situation where side equations have been used effectively is in
constructing soliton solutions to the lattice equations
\cite{AHN07,AHN08,HZ-PartII}.

Originally it was assumed that on all faces of the cube the equations
were the same in form, depending on the relevant corner variables (one
component at each corner) and spectral parameters. Recently, 3D
consistent sets have appeared, with different equations on the faces
\cite{ABS-FAA,Atk-BT,XP,LY-Ufa,Boll}. In this more general context it
is of interest to take a closer look at the Lax pairs, B\"acklund
transforms and consistency, and investigate what they are good for.
It should be noted that in the context of partial differential
equations the existence of trivial Lax pairs is well known
\cite{CN-galore} and similar examples have also been noted for some
discrete equations (see \cite{HayThesis}, Chapt.~6).

We will first recall how consistency around the cube, existence of Lax
pairs and B\"acklund transformations are intimately related for
lattice maps on the square lattice given by multi-affine relations
(sections \ref{Sec-3D-reminder} and \ref{Sec-Lax-BT}). These
considerations apply to the elementary cells, and are {\em local}.  In
section \ref{Sec-Examples}, we describe specific examples having the
CAC property. We detail in particular the already known explicit
forms of the equations and Lax pairs which we use in the rest of this
paper. In particular, we find that in some cases the zero curvature
condition (ZCC) yields two different equations that can be used to
define rational evolution in the lattice.  We then address the {\em
  global} problem of defining equations over the whole lattice, with
the guideline given by the 3 dimensional structure coming from the CAC
construction (section \ref{SecAlgE}), and check integrability with the
calculation of algebraic entropy.  In section \ref{Sec-2x2}, we push
the use of $2 \times 2$-matrix Lax pairs to its limits, by
constructing discrete systems over a larger sub-lattice of the original
lattice. This brings to the light interesting (integrable) structures
related to a generalized form of the functional Yang-Baxter
equations~\cite{HV11bis}.

\section{3D consistency, a reminder \label{Sec-3D-reminder}}
The starting point is a regular 2D square lattice, with vertices
labeled by integers $n,m$. Functions $x_{n,m}$ are associated to the
vertices, and they are subject to a constraint at all elementary
cells. This constraint is expressed by an equation
$Q(x_{n,m},x_{n+1,m},x_{n,m+1},x_{n+1,m+1})=0$, assumed to be
multi-affine in the four vertex variables. It should depend on all 4
vertex variables, and it should not factorize. It may also depend on
some parameters. Sometimes the parameters can be associated to
specific directions of the lattice, in which case they appear as
``spectral parameters''. To ease the notation, one usually denotes the
running value $x_{n,m}$ by $x_{n,m}=x$, and for neighboring values one
only indicates the shifts: $x_{n,m+1}=x_2$, or in 3D setting,
$x_1=x_{n+1,m,k},\, x_{113}=x_{n+2,m,k+1}$ etc.

\begin{figure}[h!]
\begin{center}
\setlength{\unitlength}{0.0007in}
\begin{picture}(3482,2813)(0,-10)
\put(450,1883){\circle*{91}}
\put(-150,1883){\makebox(0,0)[lb]{\small$x_{13}$}}
\put(1275,2708){\circle*{91}}
\put(800,2708){\makebox(0,0)[lb]{\small$x_{3}$}}
\put(3075,2708){\circle*{91}}
\put(3375,2633){\makebox(0,0)[lb]{\small$x_{23}$}}
\put(2250,83){\circle*{90}}
\put(2550,8){\makebox(0,0)[lb]{\small$x_{12}$}}
\put(1275,908){\circle*{90}}
\put(925,908){\makebox(0,0)[lb]{\small$x_{}$}}
\put(2250,1883){\circle*{91}}
\put(1750,2008){\makebox(0,0)[lb]{\small$x_{123}$}}
\put(450,83){\circle*{90}}
\put(-00,8){\makebox(0,0)[lb]{\small$x_{1}$}}
\put(3075,908){\circle*{90}}
\put(3300,833){\makebox(0,0)[lb]{\small$x_{2}$}}
\drawline(1275,2708)(3075,2708)
\drawline(1275,2708)(450,1883)
\drawline(450,1883)(450,83)
\drawline(3075,2708)(2250,1883)
\drawline(450,1883)(2250,1883)
\drawline(3075,2633)(3075,908)
\dashline{60.000}(1275,908)(450,83)
\dashline{60.000}(1275,908)(3075,908)
\drawline(2250,1883)(2250,83)
\drawline(450,83)(2250,83)
\drawline(3075,908)(2250,83)
\dashline{60.000}(1275,2633)(1275,908)
\end{picture}
\end{center}
\caption{The consistency cube \label{Fcac}}
\end{figure}
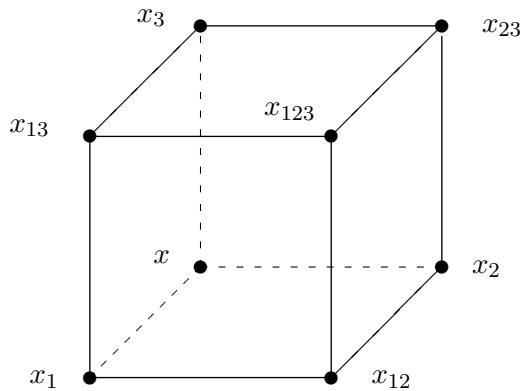

For multidimensional consistency one needs to build a cube on top of a
square and give equations on all six faces of the cube, see Figure
\ref{Fcac}, the bottom equation being the original one.

Supposing $x,x_1,x_2,x_3$ are given, one can compute $x_{12}$ using
the bottom equation, $x_{13}$ using the left side equation, and
$x_{23}$ using the back side equation.  We may then calculate the
final value $x_{123}$ in three different ways, using the front-,
right- and top-equations, respectively. The equations are said to be
Consistent-Around-a-Cube (CAC) if the three ways yield the same value
for $x_{123}$.

Note that it is also possible to check CAC with other initial values
that allow full evolution, for example, $x_3,x,x_1,x_{12}$.

\section{Lax $\simeq$ B\"acklund} \label{Sec-Lax-BT}
The Lax/BT approaches differ from CAC in that only the side equations
are used as input, while the top and bottom equations are supposed to
be derived from them.
\subsection{Construction of the Lax pair}
The Lax pair for the bottom equation is constructed from the side
equations by isolating the variables $x_3,x_{13},x_{23},x_{123}$, and
writing the equations as:
\begin{subequations}\label{side-eqs}
\begin{eqnarray}
\text{left:}\quad && x_{13} x_3 \,c_1(x_1,x)+x_{13}\, c_2(x_1,x)+x_3\,
c_3(x_1,x)+c_4(x_1,x)=0,\label{eq-left}\\ 
\text{right}\quad && x_{123} x_{23} \,\bar c_1(x_{12},x_2)+x_{123} \,
\bar c_2(x_{12},x_2)+x_{23}\, \bar c_3(x_{12},x_2)+\bar
c_4(x_{12},x_2)=0\label{eq-right},\\
\text{back}\quad && x_{23} x_3 \, b_1(x_2,x)+x_{23} \,
b_2(x_2,x)+x_3\, b_3(x_2,x)+b_4(x_2,x)=0,\label{eq-back}\\ 
\text{front}\quad && x_{123} x_{13}\, \bar b_1(x_{12},x_1)+x_{123}\,
\bar b_2(x_{12},x_1)+x_{13} \,\bar b_3(x_{12},x_1)+\bar
b_4(x_{12},x_1)=0,\label{eq-front}.
\end{eqnarray}
\end{subequations}
The coefficient functions $b_i,\bar b_i,c_i,\bar c_i$ will be  affine
linear in their arguments.

We now introduce the  homogeneous coordinates $f,g$ for $x_3$ and its
shifts by
\[
x_3=f/g,\quad x_{23}=f_2/g_2,\quad x_{13}=f_1/g_1,
\quad x_{123}=f_{12}/g_{12}.
\]
This amounts to considering $x_3$ as belonging to the projective space $
CP^1$. We denote by $\psi$ the pair
\begin{eqnarray*}
\psi = {f \choose  g}
\end{eqnarray*}
which is defined up to a global factor. Equations \eqref{side-eqs}
can then be written as
\[
\psi_1\simeq  L \psi,\quad
(\psi_{2})_1\simeq \overline{L} \psi_2,\quad
\psi_2\simeq  M  \psi,\quad
(\psi_{1})_{2}\simeq \overline{M}( \psi_1),
\]
where
\[
 L(x_1,x)\simeq \begin{pmatrix} -c_3(x_1,x) & -c_4(x_1,x) \\
 c_1(x_1,x) & c_2(x_1,x) \end{pmatrix},
 M(x_2,x) \simeq \begin{pmatrix} -b_3(x_2,x) & -b_4(x_2,x) \\
 b_1(x_2,x) & b_2(x_2,x) \end{pmatrix},\quad
\]
and similarly for the bar-quantities. Since we are working in $CP^1$
all equalities are projective equalities, and as a reminder of this we
use the symbol $\simeq$ to indicate that two matrices are equivalent
if their entries are proportional, in other words, $L$ and $M$ belong
to $PGl(2,C)$.

The Lax matrices provide parallel transport of $\psi$ along the bonds of the
lattice.  The zero curvature condition means that the parallel
transport along any closed path on the lattice is trivial. It is
necessary and sufficient to ensure that taking $\psi$ from position
$(0,0)$ to $(1,1)$ via the two routes $(0,0) \rightarrow (0,1)
\rightarrow (1,1)$ and $(0,0) \rightarrow (1,0) \rightarrow (1,1)$
gives the same result:
\begin{equation}
\label{ZCC}
(\psi_{1})_{2}\simeq (\psi_{2})_1,\quad \text{ i.e., }\quad
\overline{L}(x_{12},x_2)  M(x_2,x)\simeq \overline{M}(x_{12},x_1)   L(x_1,x).
\end{equation}
This matrix relation yields three scalar equations.\footnote{One could
  do the computations in some particular representative of the
  equivalence class, e.g., by requiring the matrices to be
  uni-modular. However this is not necessary, and may in fact be
  cumbersome, if it introduces square roots.}  These equations are
written in terms of the variables $x,x_1,x_2,x_{12}$, and in order to
satisfy the ZCC we must impose a constraint on these variables. In
standard cases, this constraint will just be the bottom
equation. Indeed, in integrable cases the three equations in
\eqref{ZCC} have a common factor. It may also happen that the ZCC is
satisfied automatically, if the side equations are simple enough, or
sometimes the common factor may factorize, yielding multiple choices.
We will examine specific examples of this phenomenon below.

\subsection{Direct approach: B\"acklund transformation}
A B\"acklund transformation (BT) is generated by a set of equations
$\mathcal E(X,Y)=0$ on two sets of functions $X$ and $Y$, such that
eliminating one set of functions say $X$ (resp. $Y$) one gets an
equations $E(Y)=0$ (resp. $E'(X)=0$) on the other set (for details,
see Proposition 4.1 in \cite{XP}.)

For a 3D consistent set of equations the side equations provide a BT
between the top and bottom equations. In order to derive the bottom
equation from the side equations one proceeds as follows:
\begin{enumerate}
\item Solve $x_{13}$ from the left equation \eqref{eq-left}.
\item Solve $x_{23}$ from the back equation \eqref{eq-back}.
\item Solve $x_{123}$ from the front equation \eqref{eq-front}.
\item After this the right equation \eqref{eq-right} can be written as
  $x_3^2\mathcal R_2+x_3\mathcal R_1+\mathcal R_0=0$, where $\mathcal
  R_i$ are polynomials in $x,x_1,x_2,x_{12}$. The greatest common
  divisor (GCD) of $\mathcal R_i$ (or one of its factors if it factorizes) will
  be the bottom equation.
\end{enumerate}
One can find the following explicit forms for the $\mathcal R_i$:
\begin{eqnarray}
&&\mathcal R_2=\det\left|\begin{array}{cccc}
c_1 & 0 & \bar b_3& \bar b_1\\
0 & b_1 & \bar c_3& \bar c_1\\
c_3 & 0 & \bar b_4& \bar b_2\\
0 & b_3 & \bar c_4& \bar c_2
\end{array}\right|,\quad \mathcal R_0=\det\left|\begin{array}{cccc}
c_2 & 0 & \bar b_3& \bar b_1\\
0 & b_2 & \bar c_3& \bar c_1\\
c_4 & 0 & \bar b_4& \bar b_2\\
0 & b_4 & \bar c_4& \bar c_2
\end{array}\right|\\
&&\mathcal R_1=\det\left|\begin{array}{cccc}
c_2 & 0 & \bar b_3& \bar b_1\\
0 & b_1 & \bar c_3& \bar c_1\\
c_4 & 0 & \bar b_4& \bar b_2\\
0 & b_3 & \bar c_4& \bar c_2
\end{array}\right|+\det\left|\begin{array}{cccc}
c_1 & 0 & \bar b_3& \bar b_1\\
0 & b_2 & \bar c_3& \bar c_1\\
c_3 & 0 & \bar b_4& \bar b_2\\
0 & b_4 & \bar c_4& \bar c_2
\end{array}\right|.
\end{eqnarray}

That the Lax and the BT approaches yield the same equations can be
seen as follows: Let us denote $V:=\overline{ M} L,\, U:=\overline{ L}
M$, then the matrix elements of $V,U$ are related to $\mathcal R_i$ as
follows (here subscripts indicate the matrix element):
\begin{eqnarray}
\frac{U_{21}}{U_{11}} -\frac{V_{21}}{V_{11}} \simeq \frac{\mathcal
  R_2}{V_{11}U_{11}},
\quad
\frac{U_{22}}{U_{12}} -\frac{V_{22}}{V_{12}} &\simeq& \frac{\mathcal R_0}{V_{12}U_{12}},\\
\frac{U_{21} V_{12}}{U_{11}V_{11}}-\frac{U_{12}
  V_{21}}{U_{11}V_{11}}
+\frac{U_{22}}{U_{11}}-\frac{V_{22}}{V_{11}}
&\simeq&\frac{\mathcal R_1}{U_{11}V_{11}}.
\end{eqnarray}
Thus $U\simeq V$ iff $\mathcal R_i=0$.

Of course one can equally well use the side equations to solve for the
variables $x_1,x_2,x_{12}$ from the left-, back-, and front-equations,
after which the right-equation will be a polynomial in $x$ with
coefficient depending on $x_3,x_{13},x_{23},x_{123}$, with their GCD
yielding the top equation.

\section{Examples}\label{Sec-Examples}
\subsection{Linear side equations}\label{Linear-Sides}
As a first example we consider the case where all the side equations
are linear, i.e., the left side equation is:
\begin{equation}\label{lin-side1}
x_{13}-x_1-x_3+x=0
\end{equation}
and we have the same equation with suitable subscript changes on the
other vertical sides. One then finds that neither the Lax nor the BT
approach yields anything, The conditions are satisfied without
reference to the bottom equation.

What about CAC? It will involve both the bottom and the top equations,
so it should give some conditions. Indeed, if one tries CAC with
\eqref{lin-side1} and a completely general multi-affine bottom and top
equations, related by a shift (but with same parameters) one finds a
consistent set with the bottom equation
\begin{equation}\label{Q1-lin}
a(x-x_1)(x_2-x_{12})+b(x-x_2)(x_1-x_{12})+c(x-x_1-x_2+x_{12})+d=0.
\end{equation}
This is a combination of the lattice modified KdV equation (lmKdV) (aka Q1
in the ABS list \cite{ABS}), and the linear equation.

A similar analysis can be done starting from  the linearizable side
equation
\begin{equation}\label{lin-side2}
x_{13}x=x_1x_3.
\end{equation}
The Lax matrix for this system is diagonal and the ZCC is
automatically satisfied. From CAC analysis we find that
\eqref{lin-side2} is compatible with the six parameter family of
homogeneous bottom/top equations of degree $2$
\begin{equation}\label{quad-lin}
{ a_1}\,x{ x_1}+{ a_2}\,x{ x_2}+{ a_3}\,x{ x_{12}}+{ a_4} \,{ x_1}\,{
  x_2}+{ a_5}\,{ x_1}\,{ x_{12}}+{ a_6}\,{ x_2} \,{ x_{12}} = 0.
\end{equation}

We will examine the integrability of these equations in section
(\ref{SecAlgE-lin}).

\subsection{H1} 

The lattice potential KdV (lpKdV), which describes the permutability
property of continuous KdV, is a paradigm of
integrable lattice equations (aka H1 in the ABS list \cite{ABS}). For
this system everything works well.  [We will return to this model in a
  new context in section~\ref{H1_2x2}.]  The model is given by
\begin{equation}\label{H1}
{\rm H1}:=(x_1-x_2)(x-x_{12})-p+q=0.
\end{equation}
After imposing this equation on the sides of the CAC cube, with
suitable variable changes, i.e.,
\begin{subequations}\label{side-H1}
\begin{eqnarray}
\text{left:}&&(x-x_{13})(x_{1}-x_3)-(p-r)=0,\\
\text{right:}&&(x_2-x_{123})(x_{12}-x_{23})-(p-r)=0,\\
\text{back:}&&(x-x_{23})(x_{2}-x_3)-(q-r)=0,\\
\text{front:}&&(x_1-x_{123})(x_{12}-x_{13})-(q-r)=0,
\end{eqnarray}
\end{subequations}
one easily finds
\begin{subequations}
\begin{eqnarray}
\mathcal R_2&=&(x_{12}-x)\cdot {\rm H1},\\ \mathcal
R_1&=&[(x-x_{12})(x_1+x_2)-\alpha-\beta+2\gamma]\cdot {\rm H1},\\ \mathcal
R_0&=&[x_1x_2(x_{12}-x)+ (\beta-\gamma)x_1+ (\alpha-\gamma)x_2]\cdot
{\rm H1},
\end{eqnarray}
\end{subequations}
with H1, as given in \eqref{H1}, as the GCD. Similarly by working on
the bottom variables $x,x_1,x_2,x_{12}$ one obtains the 3-shifted H1
as the top equation.

The Lax matrices in this case are
\begin{equation}\label{H1-l}
L(x_1,x)=\begin{pmatrix} x & p-r-x x_1\\
1 & -x_1 \end{pmatrix},\quad
M(x_2,x)=\begin{pmatrix} x & q-r-x x_2\\
1 & -x_2 \end{pmatrix},
\end{equation}
and one easily finds that
\[
 M(x_{12},x_1) \,L(x_1,x) - L(x_{12},x_2) 
\,M(x_2,x) ={\rm H1}\times\begin{pmatrix} 1 & -(x_1+ x_2)\\
0 & -1 \end{pmatrix}.
\]

\subsection{${\rm H1}_\epsilon$: a deformed version of ${\rm H1}$}
\label{H1-eps}
\begin{figure}[h!]
\begin{center}
\setlength{\unitlength}{0.0007in}
\begin{picture}(3482,2513)(0,200)
\put(450,2183){\circle*{90}}
\put(-150,2183){\makebox(0,0)[lb]{\small$x_{13}$}}

\put(1275,2708){\circle*{90}}
\put(800,2708){\makebox(0,0)[lb]{\small$x_{3}$}}
\put(3075,2708){\circle*{90}}
\put(3375,2633){\makebox(0,0)[lb]{\small$x_{23}$}}
\put(2250,383){\circle*{90}}
\put(2550,308){\makebox(0,0)[lb]{\small$x_{12}$}}
\put(1275,908){\circle*{90}}
\put(825,908){\makebox(0,0)[lb]{\small$x_{}$}}
\put(2250,2183){\circle*{90}}
\put(1650,2258){\makebox(0,0)[lb]{\small$x_{123}$}}

\put(450,383){\circle*{90}}
\put(-00,308){\makebox(0,0)[lb]{\small$x_{1}$}}

\put(3075,908){\circle*{90}}
\put(3300,833){\makebox(0,0)[lb]{\small$x_{2}$}}

\drawline(1275,2708)(3075,2708)
\drawline(1275,2708)(450,2183)

\drawline(450,2183)(450,383)
\drawline(450,2183)(2250,2183)
\drawline(450,383)(2250,383)

\drawline(3075,2708)(2250,2183)
\drawline(3075,2633)(3075,908)
\dashline{60.000}(1275,908)(450,383)
\dashline{60.000}(1275,908)(3075,908)
\drawline(2250,2183)(2250,383)
\drawline(3075,908)(2250,383)
\dashline{60.000}(1275,2633)(1275,908)
\thicklines%
{\color[rgb]{1,0,1}\drawline(1275,908)(450,2183)}%
{\color[rgb]{1,0,1}\drawline(2150,383)(2975,2708)}%
{\color[rgb]{0,1,1}\drawline(2925,2708)(1125,908)}%
{\color[rgb]{1,1,0}\drawline(2875,2708)(250,2183)}%
{\color[rgb]{1,1,0}\drawline(1025,908)(2000,383)}%
{\color[rgb]{0,1,1}\drawline(50,2183)(1850,383)}%
\end{picture}
\end{center}
\caption{The cube of H1e \label{FcacH1}}
\end{figure}
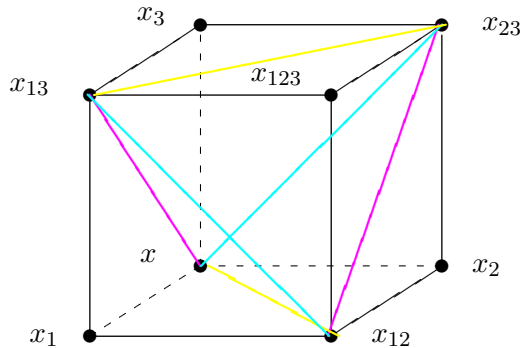

This is an asymmetric deformation of \eqref{side-H1} \cite{ABS-FAA,XP}
\begin{subequations}\label{side-H1e}
\begin{eqnarray}
\text{left:}&&(x-x_{13})(x_{1}-x_3)-(p-r)(1+\epsilon x x_{13})=0,
\\ \text{right:}&&(x_2-x_{123})(x_{12}-x_{23})-(p-r)(1+\epsilon x_{12}
x_{23})=0, \label{H1e-r}
\\ \text{back:}&&(x-x_{23})(x_{2}-x_3)-(q-r)(1+\epsilon x x_{23})=0,
\\ \text{front:}&&(x_1-x_{123})(x_{12}-x_{13})-(q-r)(1+\epsilon x_{13}
x_{12})=0.
\end{eqnarray}
The BT or Lax computations give as GCD the bottom equation
\begin{equation}
(x-x_{12})(x_{1}-x_2)-(p-q)(1+\epsilon x x_{12})=0\label{H1e-der}
\end{equation}
and similarly for the top equation we get
\begin{equation}
(x_3-x_{123})(x_{13}-x_{23})-(p-q)(1+\epsilon x_{13} x_{23})=0.
\end{equation}
\end{subequations}
If we draw a line connecting the corners appearing in the deformation
term as a product, the lines form a tetrahedron inside the cube, see
Figure \ref{FcacH1}.

The  special thing about these equations is that the parallel
sides are {\em not} obtained by shifts, but the shift must be
accompanied by a $90^o$ rotation, as described in Figure \ref{FcacH1}. 
The fact that the parallel sides are not identical
implies that the Lax matrices  associated to parallel bonds come in two
different forms. Letting
\begin{equation}\label{H1e-l}
L(x_1,x) =\begin{pmatrix} x+\epsilon (p-r)x_1 & p-r-x x_1\\
1 & -x_1 \end{pmatrix},\quad
M(x_2,x) =L|_{p\to q,\,x_1\to x_2},
\end{equation}
and 
\begin{equation}\label{H1e-lp}
L'(x_1,x) =\begin{pmatrix} x& p-r-x x_1\\
1 & -x_1 -\epsilon (p-r)x \end{pmatrix},\quad
M'(x_2,x) =L'|_{p\to q,\,x_1\to x_2},
\end{equation}
we have
\[
M'(x_{12},x_1) \,L(x_1,x) - L'(x_{12},x_2) \,M(x_2,x) =\eqref{H1e-dss}\times
\begin{pmatrix} 1-\epsilon x_1x_2 & -(x_1+ x_2)\\
-\epsilon x_1x_2 & -1+\epsilon x_1 x_2 \end{pmatrix}.
\]

Since we have two different parallel sides we can have two different
bottom equations, one derived from $M'(x_{12},x_1) \,L(x_1,x) -
L'(x_{12},x_2) \,M(x_2,x)$ as above, the other possible  choice being
\[
M(x_{12},x_1) \,L'(x_1,x) -L(x_{12},x_2) \,M'(x_2,x) =\eqref{H1e-der}\times
\begin{pmatrix} 1& -(x_1+ x_2)\\
0 & -1 \end{pmatrix}.
\]
The totality of equations on this cube, obtained from the previous one
with $L\leftrightarrow L'$ etc actually corresponds to an inversion of
the cube by
\[
x\leftrightarrow x_{123},\,x_1\leftrightarrow x_{23},\,x_2\leftrightarrow
x_{13},\, x_3\leftrightarrow x_{12},
\]
which yields
\begin{subequations}\label{side-H1e2}
\begin{eqnarray}
\text{left:}&&(x-x_{13})(x_{1}-x_3)-(p-r)(1+\epsilon x_1 x_{3})=0,\label{H1e2-l}\\
\text{right:}&&(x_2-x_{123})(x_{12}-x_{23})-(p-r)(1+\epsilon x_{2} x_{123})=0,\\
\text{back:}&&(x-x_{23})(x_{2}-x_3)-(q-r)(1+\epsilon x_2 x_{3})=0,\\
\text{front:}&&(x_1-x_{123})(x_{12}-x_{13})-(q-r)(1+\epsilon x_{1} x_{123})=0,\\
\text{bottom:}&&(x-x_{12})(x_{1}-x_2)-(p-q)(1+\epsilon x_1 x_2)=0,\label{H1e-dss}\\
\text{top:}&&(x_3-x_{123})(x_{13}-x_{23})-(p-q)(1+\epsilon x_3 x_{123})=0.
\end{eqnarray}
\end{subequations}
These are also consistent. We thus have two kinds of cubes, one defined
with \eqref{side-H1e}, and one with \eqref{side-H1e2}. These cubes can
be glued together: note that the 2-shift of \eqref{H1e2-l} is equal to
\eqref{H1e-r}. In fact under the central reflection the diagonal lines
in Figure \ref{FcacH1} would change to the other diagonals. 

We will return later to the question of gluing the cubes to fill the
space. 

\subsection{Flipped ${\rm H1}_\epsilon$\label{FH1e}}
This model was proposed in \cite{Boll} (see Sec. 3.1, Case
$(\epsilon,0,0,\epsilon)$). In that paper the model was given in a
cube with flipped coordinates $x_2\leftrightarrow x_{23}$,
$x_1\leftrightarrow x_{13}$.  In the coordinates of Figure \ref{Fcac}
the side equations of this model are given by
\begin{subequations}\label{side-H1w}
\begin{eqnarray}
\text{left:}&&(x-x_1)(x_{13}-x_3)-(p-r)(1+\epsilon x x_1)=0,\\
\text{right:}&&(x_2-x_{12})(x_{123}-x_{23})-(p-r)(1+\epsilon x_2
x_{12})=0,\\
\text{back:}&&(x-x_2)(x_{23}-x_3)-(q-r)(1+\epsilon x x_2)=0,\\
\text{front:}&&(x_1-x_{12})(x_{123}-x_{13})-(q-r)(1+\epsilon x_1
x_{12})=0.
\end{eqnarray}
\end{subequations}

\paragraph{BT derivation of bottom and top-equations.}
Now computing the values of $x_{13},x_{23},x_{123}$ from left-, back-
and right-equations, respectively, gives for the front equation an
expression that does not contain $x_3$ at all. This expression
factorizes into {\em two} factors and thus we could have two different bottom-equations:
\begin{subequations}\label{H1w-bot}
\begin{eqnarray}
\text{bottom1}&&\epsilon x x_1x_2x_{12}(1/x-1/x_1-1/x_2+1/x_{12})
-x+x_1+x_2-x_{12}=0,\\
\text{bottom2}&&p(x-x_2)(x_1-x_{12})+q(x-x_1)(x_{12}-x_2)+r(x-x_{12})(x_2-x_1)=0.\end{eqnarray}
\end{subequations}
Similarly, working with the values at the bottom square we get two
candidates for the top equations:
\begin{subequations}\label{H1w-top}
\begin{eqnarray}
\text{top1}&&x_3-x_{13}-x_{23}+x_{123}=0,\\
\text{top2}&&p(x_3-x_{23})(x_{13}-x_{123})+q(x_3-x_{13})(x_{123}-x_{23})
+r(x_3-x_{123})(x_{23}-x_{13})\nonumber\\
&&+\epsilon(p-q)(q-r)(r-p)=0.
\end{eqnarray}
\end{subequations}

Notice that in down2 and top2, there is an explicit dependence on $r$
even though these are 2D equations. In fact, one should now consider $r$
as a global parameter, although it was associated with the third
dimension in the above derivation.

Performing the usual CAC computations with the given sides and
different top and bottom equations reveals that the set of equations
is consistent in two cases: with the pair (bottom1,top1), or with
(bottom2,top2), which was given in \cite{ABS-FAA}. In other words,
given the side equations \eqref{side-H1w} there are two consistent
ways of completing the cube. Continuing further, we can also interpret
the cube to provide a BT between the left and right equations. Indeed, if
we do the above BT construction on the corners of the left side
equation the result is identically zero with (bottom1,top1), while
(bottom2,top2) produces the right-equation.

\paragraph{The Lax matrices.}
The standard procedure gives
\begin{equation}\label{H1et-l}
L(x_1,x)=\begin{pmatrix} 1 & \lambda(p,x,x_1)\\
0 & 1 \end{pmatrix},\quad
M(x_2,x)=\begin{pmatrix} 1 & \lambda(q,x,x_2)\\
0 & 1 \end{pmatrix},\quad
\lambda(a,x,y):=\frac{(a-r)(1+\epsilon x y)}{x-y}
\end{equation}
Since the matrices are upper triangular the ZCC implies
\begin{equation}\label{H1ef-lam}
\Sigma:=\lambda(p,x_2,x_{12})+\lambda(q,x,x_{2})-
\lambda(q,x_1,x_{12})-\lambda(p,x,x_{1}) = 0.
\end{equation}
Remarkably enough, the above sum factorize as
\[
\Sigma={\rm bottom1  \cdot bottom2}\,/[(x-x_1)(x-x_2)(x_1-x_{12})(x_2-x_{12})].
\]
Note that we can write \eqref{H1ef-lam} also in the form
\[
\Sigma=(T-1)\lambda(p,x,x_1)-(S-1)\lambda(q,x,x_2)=0,
\]
where $T$ is a shift in $m$ and $S$ a shift in $n$. This of course is
in the form of a conservation law.

\section{Filling the space with consistent cubes\label{SecAlgE}}
So far we have only considered a single cube and its CAC/Lax/BT. But
as the name indicates, lattice equations should be defined over the
whole lattice. This brings further complications, for example with one
cube we could freely do different M\"obius transformations in
each corner of a cube, but when the cube is part of lattice such
seemingly innocuous actions will affect neighboring cubes as well and
can destroy the lattice structure.

The rule is simple: cover the two dimensional lattice with consistent
cubes, with the condition that adjacent vertical faces coincide
exactly, that is to say their four corner values satisfy the same
equation. This is expected to produce integrable lattice equations.

We will follow this guideline for various models mentioned before, and
systematically check integrability of the lattice equations so obtained,
by calculating their algebraic entropy~\cite{AE1,AE2}: the vanishing of the
entropy is a yes/no test which gives a clear cut separation between
integrable and non integrable cases.

We may briefly recall how to calculate the entropy. The local equation
determines an evolution, starting from initial conditions given for
example on a diagonal staircase (lattice points of coordinates $(m,n)$
with $m+n=0$ or $1$). The solution is then calculated on diagonals
moving away from the diagonal of initial conditions, explicitly in
terms of these initial conditions. The algebraic entropy is defined as
the rate of growth of the degrees on these diagonals.  Exponential
growth is generic, and polynomial growth is characteristic of
integrability, while linear growth is associated with linearizable
equations~\cite{AE1,AE2,TGR}.

The exact shape of the diagonal line on which the initial values are
given is not important. If one modifies this shape locally, the
sequence of degrees will change, but not its asymptotic rate of
growth.  This should be kept in mind for some of the models studied
below. For example one could very well change the initial diagonal
(steps of height 1 and width 1) to a diagonal with bigger steps, for
example with height 2 and width 2, as this will not affect the
calculation of the entropy.

\subsection{Equations consistent with linear sides \label{SecAlgE-lin}}
We can check the integrability of the two quad equations given in
section (\ref{Linear-Sides}).  The first one, i.e., equation
(\ref{Q1-lin}), leads to the sequence of degrees:
\begin{eqnarray*}
\{ d_n \}\quad = \quad 1,\; 2,\; 4,\; 7,\; 11,\; 16,\; 22,\; 29,\;
37,\; 46,\; 67,\; 79,\dots
\end{eqnarray*}
that is to say $d_n = 1+ ( n^2 + n )/2 $. 
This quadratic growth confirms integrability.  

\noindent{\bf Remark}: Since it is integrable, it should also be
consistent with a nontrivial set of side-equations, such that one can
produce it via Lax/BT computations, but we will leave this open.

For the second equation of section (\ref{Linear-Sides}), that is to
say the general homogeneous relation (\ref{quad-lin}) of degree $2$,
the result is different. We get the following sequence of degrees:
\begin{eqnarray*}
\{ d_n \}\quad = \quad 1,\; 2,\; 4,\; 9,\; 21,\; 50,\; 120,\; 289,
\; 697 \dots
\end{eqnarray*}
which may be fitted with the generating function 
\begin{eqnarray}
g(s) = \sum_n d_n s^n = {\frac {{s}^{2}+s-1}{ \left(1- s \right)
    \left( {s}^{2}+2\,s-1 \right) }}
\end{eqnarray}
indicating a non vanishing entropy $\epsilon = \log( 1+\sqrt{2})$,
showing non-integrability.  Thus even CAC is not sharp in this case.
It was actually shown in~\cite{HV-fact} that the simple additional
condition $a_3=a_4$, renders (\ref{lin-side2}) integrable.

\subsection{H1$\epsilon$:  Checkerboard lattice }
We have already noted that in the H1$\epsilon$ model there are two
different cubes related by inversion, and that these cubes can be
glued together in a unique way, providing a black-white lattice. This
problem has been discussed in detail in \cite{XP}.

Since the gluing process is completely fixed and each cube has the CAC
property it is expected that the composite lattice is integrable.
We have calculated the sequence of degrees $\{ d_n \}$ for the
evolution defined by this model, with initial conditions on a diagonal
as prescribed above.  The outcome is the sequence
\begin{eqnarray*}
\{ d_n \} = 1 , \; 5, \; 13, \; 25, \; 41, \; 61, \; 85, \; 113, \;
145, \; 181, \; 221 ,\dots
\end{eqnarray*}
fitted by the generating function
\begin{eqnarray*}
g(s) = \sum_n d_n s^n = {\frac { \left(1+ s \right) ^{2}}{ \left( 1-s \right) ^{3}}}
\end{eqnarray*}
The above sequence has quadratic growth. Thus the entropy is vanishing
and the model is integrable.

\subsection{Flipped H1$\epsilon$: More  black and white lattices}
In the flipped H1$\epsilon$ case (Sec. \ref{FH1e}), the side equations
are the same, allowing simple gluing together of the cubes, but, as we
saw in section (\ref{FH1e}), the side-equations are somewhat weak and
allow two different compatible pairs of bottom/top equations.  We may
then cover the two dimensional lattice with consistent cubes,
assigning either equation bottom1 (top1), which we call {\em white},
or equation bottom2(top2) which we call {\em black} to each elementary
cell.  This can be done in an arbitrary way if one just insists on
having a compatible 3D structure of cubes over the 2D lattice.  It is
then natural to ask which of the configurations obtained in this way
are integrable.

Let us consider periodic distributions. The lattice is divided into
rectangular groups of cells of width $h$ and height $v$.  Within such
a rectangle, a fixed assignment is made, and the pattern is repeated
periodically in both directions.  (A pattern with $v=1$ and $h=1$
gives a uni-colored assignment.)

\begin{figure}[h!]
\begin{center} 
\includegraphics[height=2.0cm]{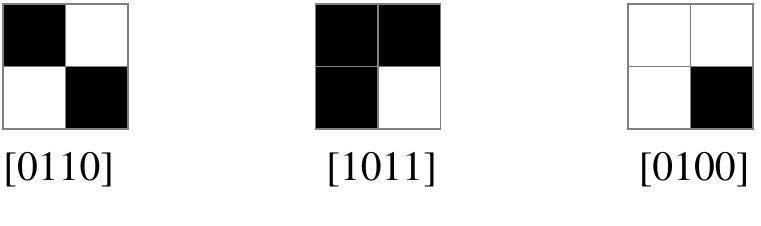}
\caption{The three inequivalent $(2,2)$ patterns.\label{Fpatt1}}
\end{center}
\end{figure}

Consider for example $(h,v)=(2,2)$. There are a priori $2^{4}$
possible patterns of that size, but only 3 inequivalent ones which
cannot be reduced to configurations having smaller periods (see Figure
\ref{Fpatt1}).  The naming convention is to list the colors starting
from the lower left corner onwards, denoting bottom1/top1 with 0,
alias white, bottom2/top2 with 1, alias black.  The equivalence of
patterns comes from the fact that we have to look at the lattice
globally. It is easy to see, for example, that in the case
$(h,v)=(2,2)$ we have the equivalences $[0100] \simeq [0010] \simeq
[1000] \simeq [0001]$, $ [1011] \simeq [1101] \simeq [0111] \simeq
[1110] $, and $ [1010] \simeq [0101] $ (checkerboard lattice).
Moreover $[0000]$ and $[1111]$ have periods $(1,1)$, $[0101]$ and
$[1010]$, $[0011]$, $[1100]$ have periods $(2,1)$ and $(1,2)$.
\bigskip
 
\noindent{\bf Claim:} {\em Some of the distributions are integrable,
  and some are not. } Although the pattern is 3D consistent, the Lax pair
is weak and cannot precisely fix the bottom and top equations.

\paragraph{$1 \times 1$ patterns (unicolor distributions).} 
Both unicolor distributions $(h,v)=(1,1)$ have vanishing entropy.  The
purely white one is linear. The purely black one is non-trivially
integrable, showing quadratic growth of the sequence of degrees
\begin{eqnarray} \label{b_seq} \{d_n\}=
 1 ,\; 2,\; 4,\; 7,\; 11,\; 16,\; 22,\; 29,\; 37,\; 46,\; 56,\; 67,\;
 79,\; 92,\; 106,\; 121,\; 137,\; 154,\; 172 \dots
\end{eqnarray}

\paragraph{$1 \times 2$ patterns.} 
Both $1 \times 2$ patterns $(h,v)=(1,2)$ or $(h,v)=(2,1)$, that is to
  say alternating black and white stripes, are integrable, with
  quadratic growth of the degrees.

\paragraph{$2 \times 2$ patterns.} 
For $(h,v)=(2,2)$ we have different results for the different patterns
in Figure \ref{Fpatt1}. 
\begin{itemize}
\item Both $[1010]$ and $[0100]$ are integrable, with quadratic growth of
the degrees.  

\item The calculation of the degrees for $[1011]$ yields the
sequence
\begin{eqnarray} 
\{d_n\}=1 ,\; 2,\; 4,\; 8,\; 18,\; 41,\; 93,\; 215,\;
493,\; 1132,\; 2600,\; 5970,\; 13710, && \nonumber\\ 31487, \;
72313,\; 166077,\; 381417,\; 875974,\; 2011788, \;
4620332  \; \dots &&  \label{1011_seq}
\end{eqnarray}
This sequence if fitted by the rational generating function
\begin{eqnarray} \label{1011_gen}
g(s) = \sum_n d_n s^n =
{\frac {1 -{s}^{2}-2\,{s}^{3}+{s}^{5}-{s}^{6}-{s}^{8}+{s}^{9}}{
 \left( 1-s \right)  \left( s+1 \right)  \left( {s}^{4}-2\,{s}^{3}-2\,
s+1 \right)  \left( {s}^{4}+1 \right) }},
\end{eqnarray}
and gives a non vanishing entropy $ \epsilon = \log(s) $ with $s$ the
largest root of ${s}^{4}-2\,{s}^{3}-2\,s+1$, approximately
$\epsilon=\log(2.29663)$.
\end{itemize}

\noindent{\bf Caveat:} When computing sequences of degrees, one should
in principle consider iterations of the whole pattern, but that tends
to make the calculations heavier. For the sequence (\ref{1011_seq}),
this would mean considering only the subsequence formed by odd terms,
leading to a growth given by the maximal root $\tau$ of
${t}^{4}-4\,{t}^{3}-6\,{t}^{2}-4\,t+1$. Of course $\tau=\sigma^2$.

\begin{figure}[h!]
\begin{center} 
\includegraphics[height=7cm]{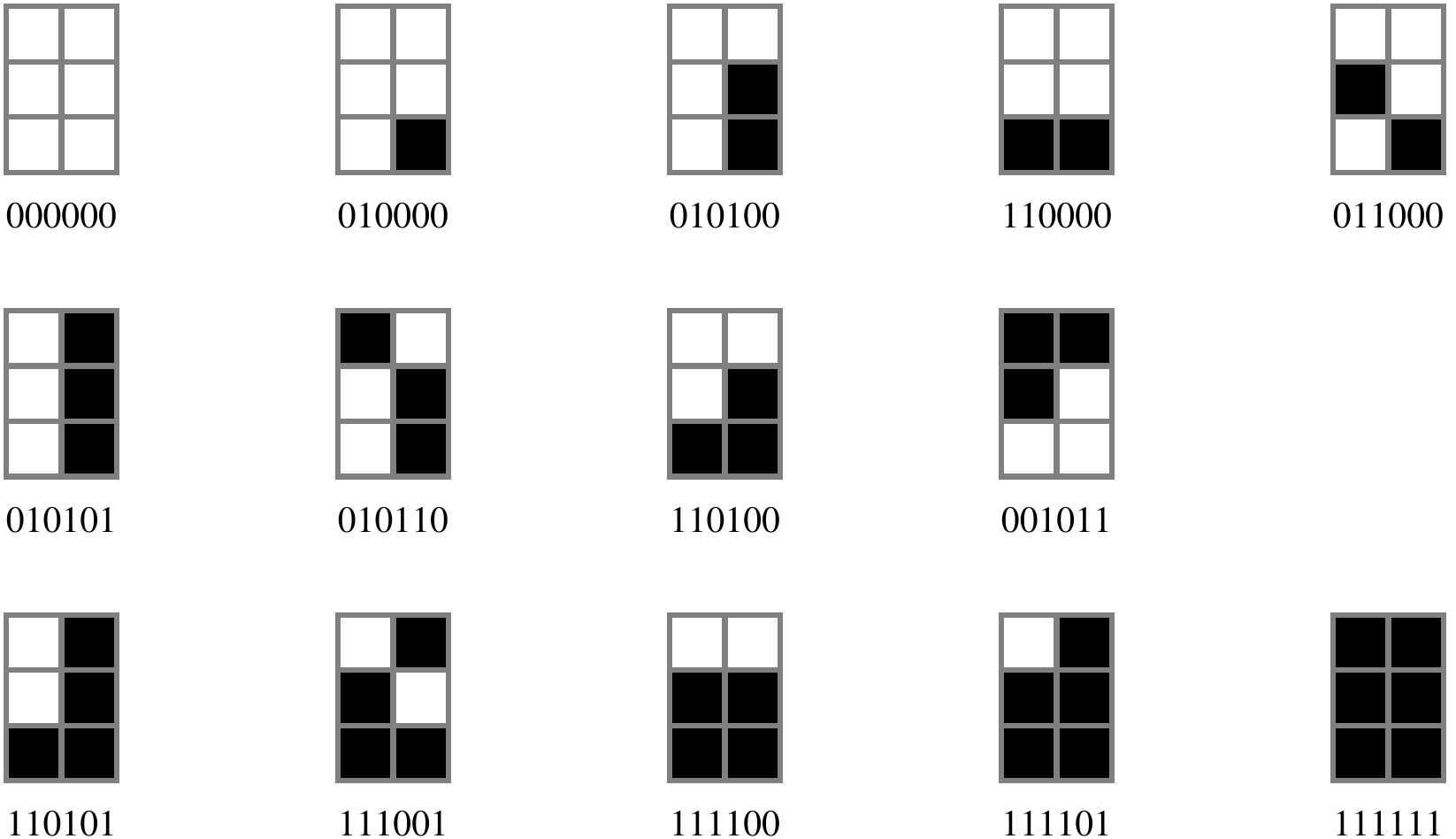}
\end{center}
\caption{Nonequivalent period $(2,3)$ patterns.\label{Fpatt23}}
\end{figure}

\paragraph{$2\times 3$ patterns.} 
We have examined all the period $(2,3)$ patterns. The various
nonequivalent patterns are depicted in Figure \ref{Fpatt23}.

The computations show that what matters is not just the proportion of
black and white cells, but the actual conformation of the pattern.
For example the period $(2,3)$ patterns $[010110]$ and $[001011]$ have
an equal number of black and white cells.  The first one is integrable
(quadratic growth of the degrees) while the latter is not, as may be
seen from the sequence:
\begin{eqnarray}\label{001011_seq } 
 \{d_n\}=
1 ,\; 1,\; 2,\; 3,\; 5,\; 9,\; 19,\; 41,\; 84,\; 169,\; 329,\; 631,\;
1199,\; 2287,\; 4412,\; 8627,\; 17059,\; 33941,\; && \nonumber
\\ 67573, \; 134071,\; 264576 ,\; 519343,\; 1015531,\; 1982461,\;
3871597, \; 7574863,\;14855790  \dots&&
\end{eqnarray}
This sequence has exponential growth, but is not long enough to
determine an exact value of the entropy. The approximate value is
$\log(1.96)$.

Out of the $14$ period $(2,3)$ nonequivalent patterns, we have one
linear case (all white $[000000]$), eight integrable cases ( $
[010000]$, $[010100]$, $[110000]$, $[011000]$, $[010101]$, $[010110]$,
$[111100]$, $[111111]$), and four non integrable ones ($[110100]$,
$[001011]$, $[110101]$, $[111001]$). The following pictures show the
aspect of two integrable cases and two non integrable ones.

\begin{figure}[h!]
\begin{center}
\includegraphics[height=3.2cm]{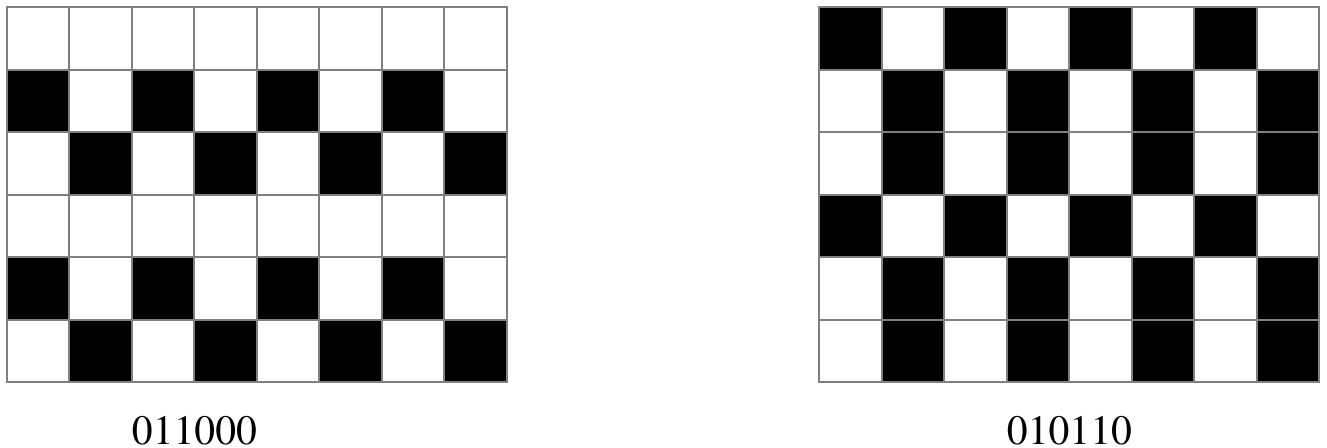}
\caption{Two integrable  $(2,3)$ patterns.}
\end{center}
\end{figure}

\begin{figure}[h!]
\begin{center} 
\includegraphics[height=3.2cm]{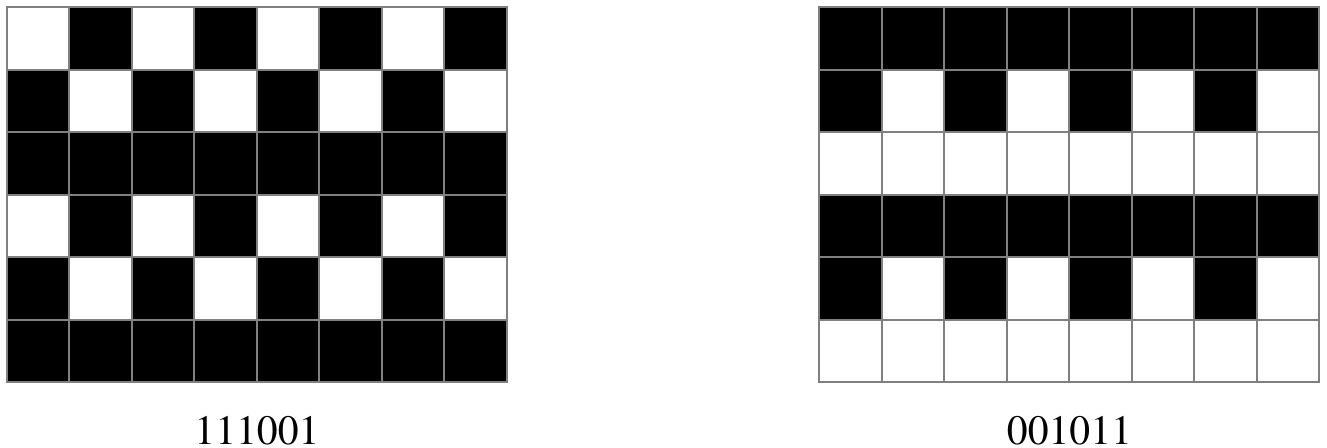}
\caption{Two nonintegrable  $(2,3)$ patterns.}
\end{center}
\end{figure}

\noindent{\bf Remark:} The entropy calculations for these patterns can
be made equally well with the relations \{ white = bottom1,
black=bottom2 \} or with \{ white = top1, black=top2\}. Both would
give the same results.

From the above results, one may already conclude that random
distributions are expected to be non-integrable.

\section{Lax pair for a $2\times 2$ sublattice}\label{Sec-2x2}
\begin{figure}[h!]
\begin{center}
\setlength{\unitlength}{0.00075in}
\begin{picture}(3482,2513)(0,-10)
\put(0,2000){\circle*{90}}
\put(2000,0){\circle*{90}}
\put(1000,0){\circle*{90}}
\put(0,1000){\circle*{90}}
\put(0,0){\circle*{90}}
\put(1000,2000){\circle{90}}
\put(2000,1000){\circle{90}}
\put(2000,2000){\circle{90}}

\drawline(0,2000)(0,0)
\drawline(1000,2000)(1000,0)
\drawline(0,2000)(2000,2000)
\drawline(2000,2000)(2000,0)
\drawline(0,0)(2000,0)
\drawline(0,1000)(2000,1000)

\put(100,40){\makebox(0,0)[lb]{$L(x_{\scriptscriptstyle 1},x)$}}
\put(1050,40){\makebox(0,0)[lb]{$ L(x_{\scriptscriptstyle 11},x_1)$}}
\put(-50,2090){\makebox(0,0)[lb]{${L(x_{\scriptscriptstyle 122},x_{\scriptscriptstyle 22})}$}}
\put(1000,2090){\makebox(0,0)[lb]{${{ L(x_{\scriptscriptstyle 1122},x_{\scriptscriptstyle 122})}}$}}

\put(-1000,450){\makebox(0,0)[lb]{$M(x_{\scriptscriptstyle 2},x)$}}
\put(2050,450){\makebox(0,0)[lb]{${{M(x_{\scriptscriptstyle 112},x_{\scriptscriptstyle 11})}}$}}

\put(-1000,1450){\makebox(0,0)[lb]{${M(x_{\scriptscriptstyle 22},x_{\scriptscriptstyle 2})}$}}
\put(2050,1450){\makebox(0,0)[lb]{${{{M(x_{\scriptscriptstyle 1122},x_{\scriptscriptstyle 112})}}}$}}

\end{picture}
\end{center}
\caption{The $2\times 2$ configuration. Values at black discs are
  initial data $(x,x_1,x_{11},x_2,x_{22})$, and values at open circles
  $(x_{112}, x_{122}, x_{1122})$ should be determined by the
  evolution.  \label{FL2x2}}
\end{figure}
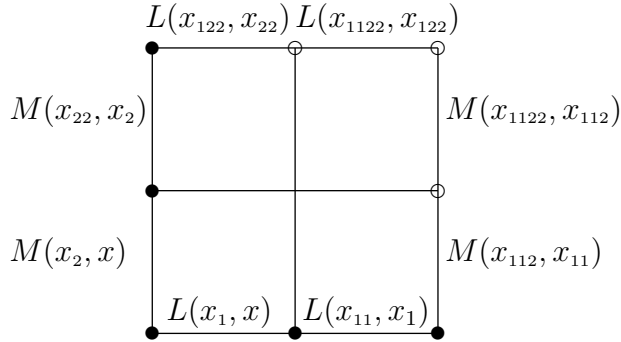

We will next push the Lax concept and ZCC to a $2\times 2$ sublattice
described in Figure \ref{FL2x2}.  (Such sub-lattices have been
discussed previously, e.g., in \cite{XP}.) To determine the evolution
we need 5 initial values, marked with black disks and expect to get
values for the vertices at the open circles. Since the Lax matrices
belong to $PGL(2,C)$ the zero curvature conditions can provide at most
three equations, and thus if everything works well the evolution is
determined. It is also clear that this will not work for bigger
sub-lattices as we would then need to provide more than 3 values.

The zero curvature condition for the $2\times 2$ sublattice is given
by
\begin{align}\label{2x2lax}
M(x_{1122},&x_{112})\, M(x_{112},x_{11})\, L(x_{11},x_1)\, L(x_1,x)\simeq\nonumber\\
&L(x_{1122},x_{122})\, L(x_{122},x_{22})\ M(x_{22},x_2)\, M(x_2,x).
\end{align}

\subsection{Flipped H1$\epsilon$}
In the flipped H1$\epsilon$ case the Lax matrices were given in
Eq.~\eqref{H1et-l} Now using this on a $2\times2$ sub-lattice we get
just one condition, namely
\begin{eqnarray*}
&&\lambda(p,x_{122},x_{1122})+ \lambda(p,x_{22},x_{122})+ 
\lambda(q,x_{2},x_{22})+ \lambda(q,x,x_{2})=\\&& 
\lambda(q,x_{112},x_{1122})+ \lambda(q,x_{11},x_{112})+ 
\lambda(p,x_{1},x_{11})+ \lambda(p,x,x_{1}).
\end{eqnarray*}
From this one can in principle solve $x_{1122}$ in terms of the other
variables. However, this equation does not determine the values for
$x_{122}$ or $x_{112}$ and therefore these Lax matrices fail to give
the evolution.

\subsection{H1} \label{H1_2x2}
For this basic model the Lax matrices were given in
Eq.~\eqref{H1-l}. Condition \eqref{2x2lax} leads to equations that
have {\em two rational} solutions: The regular one
\begin{subequations}\label{H1reg}
\begin{eqnarray}
x_{112}&=&x_1+\frac{(p-q)(x_1-x_2)}{(x_1-x_2)(x-x_{11})-(p-q)},\\
x_{122}&=&x_2+\frac{(q-p)(x_2-x_1)}{(x_2-x_1)(x-x_{22})-(q-p)},\\
x_{1122}&=&x+(p-q)\frac{(p-q)(x_{11}+x_{22}-2x)+2(x-x_{11})(x-x_{22})(x_1-x_2)}
{(p-q)^2-(x-x_{11})(x-x_{22})(x_1-x_2)^2},
\end{eqnarray}
\end{subequations}
 and an exotic solution
\begin{subequations}\label{H1ex}
\begin{eqnarray}
x_{1122}&=&x_{11}+x_{22}-x,\\
x_{122}&=&x_{112}-x_1+x_2,\\
x_{112}&=&x_1-\frac{(x_1-x_2)[(p-r)(x-x_{22})+(q-r)(x-x_{11})]}
{(p-r)(x-x_{22})-(q-r)(x-x_{11})-(x-x_{11})(x-x_{22})(x_1-x_2)},
\end{eqnarray}
\end{subequations}
The regular solution could also be obtained using the evolution on the
original lattice, first solving for $x_{12}$. As a consequence
$x_{122}$ depends only on $x,x_1,x_2,x_{22}$ and $x_{112}$ only on
$x,x_1,x_2,x_{11}$. The exotic solution is different, as $x_{122}$ and
$x_{112}$ both depend on {\em all} initial values. Furthermore, it
depends on $p-r$ and $q-r$, and not solely on $p-q$, as is the case
for the regular solution.

We may view the variables $x,x_{11},x_{22},x_{1122}$ as associated to the vertices
and  $x_1,x_2,x_{112},x_{122}$ as associated to the bonds of the
$2\times2$ sublattice.

In the algebraic entropy analysis the vertex variables are linear and
for the bond variables we find the sequence of degrees
\begin{eqnarray}
\{ d_n \} = 1,\;   4,\;   13,\;  28,\;  49,\;  76,\;  109,\; 148,\; 193,\; 244 ,   301,\;  364, 433, \dots
\end{eqnarray}
This sequence can be fitted  with the generating function
\begin{eqnarray}
\zeta(s) = \sum_n d_n s^n = {\frac {1+4\,{s}^{2}+s}{ \left(1 - s \right) ^{3}}}
\end{eqnarray}
The sequence has quadratic growth, signaling integrability.

What it the nature of the exotic solution? Since the vertex variables
have independent linear evolution we can solve the equation with
$x_{2n,2m}=F(n)+G(m)$.  When this is substituted into the bond
equations they give a non-autonomous generalization of a Yang-Baxter
map: using the coarse grained indexing $w_{n,m}=x_{2n,2m}$,
$X_{n,m}=x_{2n+1,2m},\, Y_{n,m}=x_{2n,2m+1}$, i.e.,
$x_1=X,\,x_2=Y,\,x_{112}=Y_1,x_{122}=X_2$ we have
\begin{subequations}\label{YB-FV}
\begin{equation}\label{YB-V1}
Y_1-X=P(X,Y),\quad X_2-Y=P(X,Y).
\end{equation}
The solution $w=$ constant is not allowed and if either $F$ or $G$ is
constant, $P$ collapses to $P=\pm(x-y)$. In the generic case, denoting
\[
f(n):=\frac{p-r}{F(n)-F(n+1)}, \quad g(m):=\frac{q-r}{G(m)-G(m+1)},
\]
we get
\[
P=\frac{(X-Y)[f(n)+g(m)]}{X-Y-f(n)+g(m)}.
\]
After the further translation $X\mapsto X+f(n)+T(n,m),\, Y\mapsto
Y+g(n)+T(n,m)$, where $T$ is a solution of $ T(n,m+1)-T(n,m)=2g(m),\,
T(n+1,m)-T(n,m)=2f(n)$
we finally get
\begin{equation}\label{YB_V2}
P=\frac{f(n)^2-g(m)^2}{X-Y},
\end{equation}
\end{subequations}
which is a non-autonomous version of the Adler map
\cite{AdMap} (aka $F_V$ in the classification \cite{ABS-CAG}).  The
situation can be described by the following diagram:
\[\begin{CD}
\text{standard solution} @<<<\text{H1 Lax for }2\times2\text{ sublattice} @>>> \text{exotic solution }\eqref{H1ex}\\
 @. @AAA @VVV\\
@. {\rm H1 } @. \text{non-autonomous }F_V\, \eqref{YB-FV}
\end{CD}
\]

\subsection{H3} \label{H3_2x2}
The phenomenon described in the previous section is not
generic. Indeed, applying the same coarse-graining to an arbitrary
integrable quad-equation will lead to a system having only one
rational solution (the regular one coming from the original lattice).

We have, however, found more examples where an exotic rational
solution exists. Here is one, provided by the lattice modified KdV
(lmKdV) (aka H3$_{\delta=0}$). In that case the defining relations of
the exotic solution are:
\begin{subequations}
\begin{eqnarray}
x_{1122}\, x&=&x_{11}\, x_{22},\label{H3ver}\\
x_{122}\,x_1&=&x_{112}\,x_2,\\
\frac{x_{122}}{x_2}=\frac{x_{112}}{x_1}&=&
\frac{(q^2 x_{22}-r^2 x)(x-x_{11})p x_1-(p^2 x_{11}-r^2 x)(x-x_{22})q x_2}
{(r^2x_{22}-q^2 x)(x_{11}-x)p x_2-
(r^2x_{11}-p^2 x)(x_{22}-x)q x_1}
\end{eqnarray}
\end{subequations}
In the algebraic entropy analysis the sequence of degrees for the
vertex variables has linear growth as expected, while the sequence for the
bonds is the same as for H1 (see above).

Now the equation on the vertex variables \eqref{H3ver} can be solved with
\[
x_{2n,2m}=F(n)G(m),
\]
and if we introduce 
\[
f(n):=\frac{r^2F(n+1)-p^2F(n)}{F(n+1)-F(n)},\quad
g(m):=\frac{r^2G(m+1)-q^2G(m)}{G(m+1)-G(m)},
\]
we obtain the bond equations in the form
\begin{equation}\label{H3int}
\frac{X_2}{Y}=\frac{Y_1}{X}=
\frac{(q^2+r^2-g(m))p X-(p^2+r^2-f(n))q Y}{f(n)q X-g(m)p Y},
\end{equation}
using the previously introduced notation. With the further scaling
\[X(n,m)\mapsto p\,T(n,m)\,X(n,m)/f(n),
Y(n,m)\mapsto q\,T(n,m)\,Y(n,m)/g(m),
\]
 where $T$ solves
\[
T(n,m+1)/T(n,m)=(q^2+r^2-g(m))/g(m),\quad
T(n+1,m)/T(n,m)=(p^2+r^2-f(n))/f(n),
\]
equation \eqref{H3int} reduces to
\begin{equation}
X_2 =\frac{Y}{\alpha(n)}\, P,\quad 
Y_1 =\frac{X}{\beta(m)}\, P,\quad 
P=\frac{\alpha(n)\, X-\beta(m) Y}{X-Y},
\end{equation}
where
\[
\alpha(n)=\lambda p^2/[f(n)(f(n)-p^2-r^2)],\quad
\beta(n)=\lambda q^2/[g(n)(g(n)-q^2-r^2)].
\]
This is nothing but a non-autonomous version of $F_{III}$ in the
classification of \cite{ABS-CAG}. The diagram presented for H1
works also for H3$(\delta=0)$.

We have found that this phenomenon occurs also for the lattice
modified KdV and for the lattice Schwarzian KdV.  We have examined in
some detail the properties of the models defined in this way and we
will present these results elsewhere \cite{HV11bis}.

\section{Discussion}
We have discussed the strength of the Lax pairs (or the zero curvature
condition) and BT through various examples. We have found several
cases where the ZCC does not uniquely determine the evolution but
allows two possibilities. This happens, e.g., in the flipped
H1$\epsilon$ model. If one then builds an infinite lattice by
arbitrarily choosing for each cell one of the two allowed relations,
the result is sometimes integrable and sometimes shows nonzero
entropy.

If the ZCC is pushed to a $2\times2$ sublattice we get more examples
where it is ambiguous and yields both the regular solution as well as
an exotic one. The latter cannot be generated by some equation in the
sublattice, because in the exotic solution the variables $x_{122}$ and
$x_{112}$ depend on both $x_{11}$ and $x_{22}$ which is not possible
using the equations on the elementary squares.

The equations we have obtained in this way can be interpreted as having
vertex and edge variables: the variables with even number of indices
live at the vertices while the ones with an odd number of indices live
on the edges. The edge variables evolve as in a non-autonomous 
functional Yang-Baxter equation. For more details see \cite{HV11bis}.

\subsection*{Acknowledgments}
One of us (JH) was partially supported by the Ville de Paris through
the program ``Research in Paris 2010''. Some of the computations were done
using REDUCE \cite{Red}.


\begin{thebibliography}{99}
\bibitem{NRGO} FW Nijhoff, A Ramani, B Grammaticos and Y Ohta, {\it On
  Discrete Painlev\'e Equations Associated with the Lattice KdV
  Systems and the Painlev\'e VI Equation},
    Stud. Appl. Math. {\bf 106} (2001) 261.

\bibitem{NW} FW. Nijhoff and A. Walker, {\em The discrete and
    continuous Painlev\'e VI hierarchy and the Garnier systems},
  Glasgow Math. J. {\bf 43A}, 109 (2001).

\bibitem{BS-IMR} A.I. Bobenko, Yu.B. Suris. {\em Integrable systems on
quad-graphs.}  Internat. Math. Res. Notices, {\bf 11},
573-611 (2002).

\bibitem{ABS} V Adler, A Bobenko and Yu Suris, {\em Classification of
  Integrable Equations on Quad-Graphs.  The Consistency Approach},
  Commun. Math.  Phys. {\bf 233} (2003) 513.

\bibitem{Nlax} FW. Nijhoff, {\em Lax pair for the Adler (lattice
  Krichever--Novikov) system}, Phys. Lett. {\bf A297} 49-58 (2002).

\bibitem{AHN07} J. Atkinson, J. Hietarinta and FW. Nijhoff, {\it Seed
  and soliton solutions of Adler's lattice equation}, J.Phys.A:
  Math. Theor.  {\bf 40} (2007) F1--F8.

\bibitem{AHN08} J. Atkinson, J. Hietarinta and FW. Nijhoff, {\it
  Soliton solutions for Q3}, J. Phys. A: Math. Theor.  {\bf 41}
  (2008) 142001 (11 pp).

\bibitem{HZ-PartII} J. Hietarinta and D.J. Zhang, {\em Soliton
    solutions for ABS lattice equations: II Casoratians and
    bilinearization}, J. Phys. A: Math. Theor. {\bf 42} (2009) 404006 (30pp).

\bibitem{ABS-FAA} V. E. Adler, A. I. Bobenko, and Yu. B. Suris.  {\em
  Discrete Nonlinear Hyperbolic Equations. Classification of Integrable
  Cases}, Funct. Anal. App., {\bf 43}, 3-17 (2009).

\bibitem{Atk-BT} J. Atkinson. {\em B\"acklund transformations for
  integrable lattice equations} J. Phys. A: Math. Theor. {\bf 41}
  135202  (2008).

\bibitem{XP} P D Xenitidis and V G Papageorgiou.
{\em Symmetries and integrability of discrete equations
defined on a black--white lattice}, J. Phys. A: Math. Theor. {\bf 42} (2009)
454025 (13pp.).

\bibitem{LY-Ufa} D. Levi and R. I. Yamilov. {\em On a nonlinear
  integrable difference equation on the square}, Ufimsk. Mat. Zh. {\bf
  1}, 101--105 (2009).

\bibitem{Boll} R. Boll. {\em Classification of 3D consistent
  quad-equations}, {\tt arXiv:1009.4007v2 [nlin.SI] 3 Nov 2010}.

\bibitem{CN-galore} F. Calogero and M.C. Nucci, {\em  Lax pairs galore},
J. Math. Phys. {\bf 32}, 72 (1991).

\bibitem{HayThesis} Mike Hay, {\em Discrete Lax Pairs, Reductions and
Hierarchies}, Thesis (University of Sydney, 2008).

\bibitem{HV11bis} J. Hietarinta and C. Viallet, {\em Integrable
    lattices with vertex and bond variables}.

\bibitem{AE1}  M. Bellon and C-M. Viallet. {\em   Algebraic Entropy} 
 {Comm. Math. Phys.} {\bf 204} (1999) {425--437}.

\bibitem{AE2} C-M. Viallet {\em Algebraic entropy for lattice
  equations}, {\tt arXiv:math-ph/0609043}

\bibitem{TGR} S.~Tremblay, B.~Grammaticos, and A.~Ramani, {\em
  Integrable lattice equations and their growth properties},
  Phys. Lett. {\bf A 278} (2001) 319--324.


\bibitem{HV-fact} J. Hietarinta and C. Viallet.  {\em Searching for
  integrable lattice maps using factorization}, J. Phys. A:
  Math. Theor. {\bf 40} (2007) 12629--12643

\bibitem{AdMap} V.E. Adler, {\em Recuttings of polygons},
  Funct. Anal. App., {\bf 27}, 141--143 (1993).

\bibitem{ABS-CAG} V.E. Adler, A.I. Bobenko and Yu.B. Suris, {\em
  Geometry of Yang–Baxter Maps: pencils of conics and quadrirational
  mappings}, Commun. Anal. Geom., {\bf 12}, 967-1007 (2004).

\bibitem{Red} A. Hearn, {\em REDUCE User's Manual Version
  3.8} (2004) \hfill \\ {\tt http://reduce-algebra.sourceforge.net/}


\end{thebibliography}
\end{document}